\documentstyle[12pt]{article}

\begin{document}
\setcounter{page}{1}
\pagestyle{plain} \vspace{1cm}
\begin{center}
\Large{\bf Bouncing Universe with a Nonminimally Coupled Scalar Field on a Moving Domain Wall}\\
\small \vspace{1cm}
{\bf Kourosh Nozari$^{a,b}$}\quad\ and \quad {\bf S. Davood Sadatian$^{a,c,d}$}\\
\vspace{0.5cm} {\it $^{a}$Department of Physics,
Faculty of Basic Sciences,\\
University of Mazandaran,\\
P. O. Box 47416-95447,
Babolsar, IRAN,\\
$^{b}$Research Institute for Astronomy and Astrophysics of Maragha,
\\P. O. Box 55134-441, Maragha, IRAN, \\
$^{c}$ Nishabour Center of Higher Education, Nishabour, IRAN\\
and\\
$^{d}$Islamic Azad University, Nishabour Branch, Nishabour, IRAN}\\

{\it knozari@umz.ac.ir\\
d.sadatian@umz.ac.ir}
\end{center}
\vspace{1.5cm}
\begin{abstract}
We study dynamics of a dark energy component nonminimally coupled to
gravity on a moving domain wall. We use this setup to explain
late-time accelerated expansion and crossing of the phantom divide
line by the equation of state parameter of this non-minimally
coupled dark energy component. By analyzing parameter space of the
model, we show that this model accounts for accelerated expansion
and crossing of the phantom divide line with a suitable fine-tuning
of the nonminimal coupling. Then we study the issue of bouncing
solutions in this framework.
\end{abstract}
\vspace{1.5cm}
\newpage

\section{Introduction}
Recent evidences from supernova searches data [1,2], cosmic
microwave background (CMB) results [3-5] and also Wilkinson
Microwave Anisotropy Probe (WMAP) data [6,7], indicate an positively
accelerating phase of the cosmological expansion today and this
feature shows that the simple picture of universe consisting of
pressureless fluid is not enough. In this regard, the universe may
contain some sort of additional negative-pressure component dubbed
dark energy. Analysis of the WMAP data [8-10] shows that there is no
indication for any significant deviations from Gaussianity and
adiabaticity of the CMB power spectrum and therefore suggests that
the universe is spatially flat to within the limits of observational
accuracy. Further, the combined analysis of the WMAP data with the
supernova Legacy survey (SNLS) [8], constrains the equation of state
$w_{de}$, corresponding to almost ${74\%}$ contribution of dark
energy in the currently accelerating universe, to be very close to
that of the cosmological constant value. Moreover, observations
appear to favor a dark energy equation of state, $w_{de}<-1$ [11].
Therefore, a viable cosmological model should admit a dynamical
equation of state that might have crossed the value $w_{de}= -1$, in
the recent epoch of cosmological evolution. In fact, to explain
positively accelerated expansion of the universe, there are two
alternative approaches: incorporating an additional cosmological
component or modifying gravity at cosmological scale.
Multi-component dark energy with at least one non-canonical phantom
field is a possible candidate of first alternative. This viewpoint
has been studied extensively in literature ( see [12] and references
therein ). Another alternative to explain current accelerated
expansion of the universe is extension of general relativity to more
general theories on cosmological scales. In this view point,
modified Einstein-Hilbert action resulting $f(R)$-gravity ( see [13]
and references therein) and braneworld gravity [14-16] are studied
extensively. For instance, DGP ( Dvali-Gabadadze-Porrati) braneworld
scenario as an infra-red modification of general relativity explains
accelerated expansion of the universe in its self-accelerating
branch via leakage of gravity to extra dimension. In this model,
equation of state parameter of dark energy never crosses
$\omega(z)=-1$ line, and universe eventually turns out to be de
Sitter phase. Nevertheless, in this setup if we use a single scalar
field (ordinary or phantom) on the brane, we can show that equation
of state parameter of dark energy component can cross phantom divide
line [17]. Also quintessential behavior can be achieved in a
geometrical way in higher order theories of gravity [18]. One
important consequence in quintessence model is the fact that a
single minimally coupled scalar field has not the capability to
explain crossing of the phantom divide line, $\omega_{\phi}=-1$ [19,
20],(see also [21]). However, a single but non-minimally coupled
scalar field is adequate to cross the phantom divide line by its
equation of state parameter [12]. On the other hand, in the context
of scalar-vector-tensor theories, realizing accelerated expansion
and crossing of the phantom divide line with one minimally coupled
scalar field in the presence of a Lorentz invariance violating
vector field has been reported [22].

In this letter, we consider a nonminimally coupled scalar field as a
dark energy component on a moving domain wall. In this extension,
brane is considered as a moving domain wall in a background
$5$-dimensional anti de Sitter-Schwarzschild (AdSS$_{5}$) black hole
bulk. In other words, we consider a static bulk configuration with
two $5$-dimensional anti de Sitter-Schwarzschild black hole spaces
joined by a moving domain wall. Then we study dynamics of equation
of state parameter of a non-minimally coupled scalar field on this
moving domain wall. We show that adopting a phenomenologically
appropriate ansatz with suitable fine-tuning of the parameters of
the model, provide enough room to explain accelerated expansion and
crossing of the phantom divide line by the dark energy equation of
state parameter. We also investigate the existence of bouncing
solutions in this setup. Based on recent observational data,
parameters of this model are constrained in the favor of late-time
accelerated expansion. The importance of this study lies in the fact
that currently, models of phantom divide line crossing are so
important that they can realize that which model is better than the
others to describe the nature of dark energy. In this respect,
possible crossing of the phantom divide line(PDL), $\omega=-1$, by
equation of state parameter of nonminimally coupled scalar field and
existence of bouncing solutions in this braneworld setup are discussed.

\section{The Setup}
We consider a moving domain wall picture of braneworld [23-25], to
discuss the issue of quintessence and late-time acceleration along
with the phantom divide line crossing of the equation of state
parameter of a non-minimally coupled scalar field on the brane.
Following [23], we consider a static bulk configuration with two
$5$-dimensional anti de Sitter-Schwarzschild (AdSS$_{5}$) black hole
spaces joined by a moving domain wall. To embed this moving domain
wall into $5$-dimensional bulk, it is then necessary to specify
normal and tangent vectors to this domain wall with careful
determination of normal direction to the brane. We assume that
domain wall is located at coordinate $r=a(\tau)$ where $a(\tau)$ is
determined by Israel junction conditions [26]. In this model,
observers on the moving domain wall interpret their motion through
the static $5$-dimensional bulk background as cosmological expansion
or contraction. Now, consider the following line element [23]
\begin{equation}
{{dS}_{5\pm}}^{2}=-\bigg(k-\frac{\eta_{\pm}}{r^2}+\frac{r^2}{\ell^{2}}\bigg)dt^{2}
+\frac{1}{k-\frac{\eta_{\pm}}{r^2}+\frac{r^2}{\ell^{2}}}dr^{2}+r^{2}\gamma_{ij}dx^{i}dx^{j},
\end{equation}
where $\pm$ stands for left($-$) and right($+$) side of the moving
domain wall, $\ell$ is curvature radius of AdS$_{5}$ manifold and
$\gamma_{ij}$ is the horizon metric of a constant curvature manifold
with $k=-1,\, 0,\,1$ for open, flat and closed horizon geometry
respectively and $\eta_{\pm}\neq 0$ generates the electric part of
the Weyl tensor on each side. This line element shows a topological
anti de Sitter black hole geometry in each side. Using Israel
junction conditions [26] and Gauss-Codazzi equations, we find the
following generalization of the Friedmann and acceleration equations
[23]
\begin{equation}
\frac{\dot{a}^{2}}{a^2}+\frac{k}{a^{2}}=\frac{\rho}{3}+\frac{\eta}{a^{4}}+
\frac{\ell^{2}}{36}\rho^{2},
\end{equation}
\begin{equation}
\frac{\ddot{a}}{a}=-\frac{\rho}{6}(1+3w)-\frac{\eta}{a^4}-\frac{\ell^2}{36}\rho^{2}(2+3w),
\end{equation}
where we have adapted a $Z_{2}$-symmetry with
$\eta_{+}=\eta_{-}\equiv\eta$ and $\omega$ is defined as
$\omega=\frac{p}{\rho}$. We assume there is a scalar field
non-minimally coupled to gravity on the moving domain wall. The
action of this non-minimally coupled scalar field is defined as
[27,28,29]
\begin{equation}
S_{\varphi}=\int
d^{4}x\sqrt{-g}\bigg[\frac{1}{{k_{4}}^{2}}\alpha(\varphi)
R[g]-\frac{1}{2} g^{\mu\nu} \nabla_{\mu}\varphi\nabla_{\nu}\varphi
-V(\varphi) \bigg].
\end{equation}
Energy-momentum tensor of this non-minimally coupled scalar field is
given by
\begin{equation}
{\cal{T}}_{\mu\nu}=\nabla_{\mu}\varphi\nabla_{\nu}\varphi-\frac{1}{2}g_{\mu\nu}(\nabla\varphi)^{2}-
g_{\mu\nu}V(\varphi)+g_{\mu\nu}\Box\alpha(\varphi)-\nabla_{\mu}\nabla_{\nu}\alpha(\varphi),
\end{equation}
where $\Box$ shows $4$-dimensional d'Alembertian. We assume a FRW
type universe on the brane with line element defined as
\begin{equation}
ds^{2}=-dt^{2}+a^{2}(t)d{\Sigma_{k}}^{2},
\end{equation}
where $d{\Sigma_{k}}^{2}$ is the line element for a manifold of
constant curvature $k = +1,0,-1$. Then equation of motion for scalar
field $\phi$ is
\begin{equation}
\nabla^{\mu}\nabla_{\mu}\varphi=V'-\alpha'R[g],
\end{equation}
where a prime denotes the derivative of any quantity with respect
to\, $\varphi$. This equation can be written as
\begin{equation}
\ddot{\varphi}+3\frac{\dot{a}}{a}\dot{\varphi}+\frac{dV}{d\varphi}=
\alpha'R[g]
\end{equation}
where a dot denotes the derivative with respect to cosmic time,\,
$t$\, and Ricci scalar is given by
\begin{equation}
R=6\bigg(\dot{H}+2H^{2}+\frac{k}{a^{2}}\bigg).
\end{equation}
This non-minimally coupled scalar field localized on the brane will
play the role of dark energy component in our setup. We assume this
field has only time dependence. The energy density and pressure of
this non-minimally coupled dark energy component are given as
follows
\begin{equation}
\rho=\alpha^{-1}\bigg(\frac{1}{2}\dot{\varphi}^{2}+V(\varphi)-6\alpha'H\dot{\varphi}\bigg),
\end{equation}
\begin{equation}
p=\alpha^{-1}\bigg[\frac{1}{2}\dot{\varphi}^{2}-V(\varphi)+
2\Big(\alpha'\ddot{\varphi}+2H\alpha'\dot{\varphi}+\alpha''\dot{\varphi}^2\Big)\bigg],
\end{equation}
where $H=\frac{\dot{a}}{a}$ is Hubble parameter on the moving domain
wall. Now, assuming that brane is tensionless, in which follows we
discuss two cases with $\eta=0$  and  $\eta\neq 0$\, separately.
Note that $\eta$ is the coefficient of a term which is called dark
radiation term. For $\eta=0$ ( the corresponding term can be
neglected in late-time due to fast decay), each sub-manifolds of
bulk spacetime are exact AdS$_{5}$ spacetimes. With a localized
non-minimally coupled scalar field as the only source of
energy-momentum on the brane, we discuss late-time acceleration and
phantom divide line crossing in this setup. For this purpose, we use
energy density and pressure of scalar field defined in equations
(10) and (11) to rewrite equation (3) with $\eta=0$ as
follows\footnote{Note that equations (10) and (11) contain
additional terms proportional to $\varphi^{2}$ which we have
neglected due to exponentially decreasing ansatz for $\varphi$ used
in this paper.}
$$\frac{\ddot{a}}{a}=-\frac{1}{6\alpha}\bigg(\frac{1}{2}\dot{\varphi}^{2}+V(\varphi)-
6\alpha'H\dot{\varphi}\bigg)\bigg(1+3\frac{\dot{\varphi}^{2}-2V(\varphi)+
4\big(\alpha'\ddot{\varphi}+2H\alpha'\dot{\varphi}+\alpha''\dot{\varphi}^2\big)}
{\dot{\varphi}^{2}+2V(\varphi)-12\alpha'H\dot{\varphi}} \bigg)$$
\begin{equation}
-\frac{\ell^2}{36\alpha^2}\bigg(\frac{1}{2}\dot{\varphi}^{2}+V(\varphi)-
6\alpha'H\dot{\varphi}\bigg)^{2}\bigg(2+3\frac{\dot{\varphi}^{2}-2V(\varphi)+
4\big(\alpha'\ddot{\varphi}+2H\alpha'\dot{\varphi}+\alpha''\dot{\varphi}^2\big)}
{\dot{\varphi}^{2}+2V(\varphi)-12\alpha'H\dot{\varphi}}\bigg),
\end{equation}
This is a complicated relation and to explain its cosmological
implications, we have to consider either some limiting cases or
specify $\alpha(\varphi)$, $V(\varphi)$ and $\varphi$. Before
further discussion, we note that due to existence of several
fine-tunable parameters and a combination of plus and minus signs in
this relation, essentially it is possible to find a domain of
parameter space that satisfies the condition $\ddot{a}>0$ in the
favor of positively accelerated expansion. Now, to proceed further,
we assume $\alpha(\varphi)=\frac{1}{2}\Big(1-\xi \varphi^{2}\Big)$
which is corresponding to conformal coupling of the scalar field and
gravity on the brane. We apply a phenomenologically reliable ansatz
( see for instance [30]) so that \,$\varphi=\varphi_{0} e^{-kt}$\,
and\, $a=(t^2+\frac{t_0}{1-\nu})^{\frac{1}{1-\nu}}$\, where $k$\,
and $t_0$ are positive constants. Here we assume $\nu\neq1$. Also we
set $V=\lambda \varphi^{n}$. Defining $A=\frac{\ddot{a}}{a}$,
equation (12) with this ansatz takes the following form
\begin{equation}
A=-{\frac {{\it E_{1}}\, \left( 1+3\,{\it E_{2}} \right)
}{3-3\,\xi\,\varphi}}-{ \frac {{\ell}^{2}{{\it E_{1}}}^{2} \left(
2+3\,{\it E_{2}} \right) }{9-9\,\xi\, \varphi}}
\end{equation}
where
$$E_{1}\equiv \frac{1}{2}{{\phi_0}}^{2}{k}^{2}\left({e^{-kt}}\right)^{2}+\lambda\left({\phi_0}{e^{-kt}}\right)^{n}-12
{\frac {\xi{{\phi_0}}^{2}\left({e^{-kt}}\right)^{2}tk}{\left(1-\nu
\right){t }^{2}+{t_0}}}
$$ and
$$E_{2}\equiv \frac{\left({{\phi_0}}^{2}{k}^{2}\left({e^{-kt}}\right)^{2}-2\lambda
\left({\phi_0}{e^{-kt}}\right)^{n}-8\xi{{\phi_0}}^{2}\left(
{e^{-kt}}\right)^{2}{k}^{2}+16{\frac{\xi{{\phi_0}}^{2}\left(
{e^{-kt}}\right)^{2}kt}{\left(1-\nu \right){t}^{2}+{t_0}}}\right)}
{\left({{\phi_0}}^{2}{k}^{2}\left({e^{-kt}}\right)^{2}+2\lambda
\left({\phi_0}{e^{-kt}}\right)^{n}-24{\frac{\xi{{\phi_0}}^{2}\left(
{e^{-kt}}\right)^{2}kt}{\left(1-\nu\right){t}^{2}+{t_0}}}\right)}$$
Figure $1$ shows the possibility of accelerated expansion ( $A>0$
for $\nu<1$ in some appropriate domain of parameter space ( for
example with $\lambda=\ell=1$\,, $\xi=0.15$\,, $k=0.1$\,and $n=2$)).
\begin{figure}[htp]
\begin{center}\includegraphics{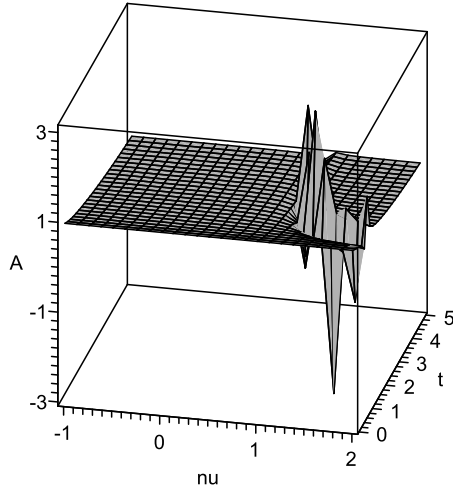} \vspace{6cm}
\end{center}
 \caption{\small {Accelerated expansion with a nonminimally coupled scalar
 field on the moving brane embedded in AdS$_{5}$ bulk.}}
\end{figure}
The case with $\eta\neq 0$ accounts for accelerated expansion in
even more simpler manner due to its wider parameter space. In figure
$1$, we see that for $\nu>1$ equation (13) has unusual behavior then
therefore we restrict ourselves to cases with $\nu<1$. In this
braneworld setup, equation of state parameter with above ansatz has
the following form
\begin{equation}
\omega=\frac{p}{\rho}=\frac{\left(1/2{{\phi_0}}^{2}{k}^{2}\left(
{e^{-kt}}\right)^{2}-\lambda\left({\phi_0}{e^{-kt}}\right)^{n}-4\xi{{\phi_0}}^{2}
\left({e^{-kt}}\right)^{2}{k}^{2}+8{\frac{t\xi{{\phi_0}}^{2}\left(
{e^{-kt}}\right)^{2}k}{\left(1-\nu \right){t}^{2}+{t_0}}}\right)
}{\left(1/2{{\phi_0}}^{2}{k}^{2}\left({e^{-kt}}\right)^{2}+\lambda
\left({\phi_0}{e ^{-kt}}\right)^{n}-12{\frac{t\xi{{\phi_0}}^{2}
\left({e^{-kt}}\right)^{2}k}{\left(1-\nu\right){t}^{2}+{t_0}}}
 \right)}
\end{equation}
Figure $2$ shows the dynamics of equation of state parameter in this
model with aforementioned ansatz. As this figure shows, equation of
state parameter of this model crosses the phantom divide,
$\omega=-1$ line. On the other hand, accelerated expansion with
$\eta\neq0$ is easily achieved due to wide parameter space of this
setup. As we have emphasized in introduction, currently models of
phantom divide line crossing are so important that they can realize
that which model is better than the others to describe the nature of
dark energy. In this sense a non-minimally coupled scalar field on
the brane provides a good candidate for explaining accelerated
expansion and crossing of the phantom divide line as a suitable
candidate of dark energy.
\begin{figure}[htp]
\begin{center}\includegraphics{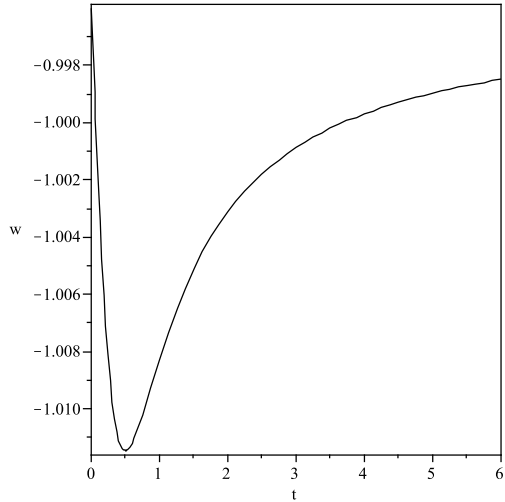} \vspace{5.5cm}
\end{center}
 \caption{\small {Crossing of the phantom divide line by equation of state parameter of
 a non-minimally coupled scalar field on the moving brane embedded
 in AdS$_{5}$ bulk. Based on analysis of [31,32]( see also [35]), we have set $\xi=0.15$
 in numerical calculation. We have set also $\nu=-3$).}}
\end{figure}

\section{Bouncing Solutions}
A possible solution of the singularity problem of the standard Big
Bang cosmology is the so-called Bouncing Universe. A bouncing
universe has an initial contracted state with a non-vanishing
minimal radius and then evolves to an expanding phase [30,33]. To
have a successful bouncing model in the framework of standard
cosmology, the null energy condition is violated for a period of
time around the bouncing point. Moreover, for the universe entering
into the hot Big Bang era after the bouncing, the equation of state
parameter of matter content of the universe must transit from
$\omega<-1$ to $\omega>-1$ ( see for instance [30]).

As figure $2$ shows, in our model equation of state parameter
crosses the $\omega=-1$ line from $\omega>-1$ to  $\omega<-1$ and
this is supported by observations [34]. This is a dynamical model of
dark energy which differs from models with just a cosmological
constant, pure quintessence field, pure phantom field, K-essence and
so on in the determination of the cosmological evolution. However,
in some sense this model is similar to quintom dark energy models
which consist two fields: one quintessence and the other phantom
field [35]. In which follows, we study necessary conditions required
for a successful bounce in a model universe dominated by the
nonminimally coupled scaler field on the moving brane embedded in
AdS$_{5}$ bulk. During the contracting phase, the scale factor
$a(t)$ is decreasing, i.e., $\dot a(t)<0$, and in the expanding
phase we have $\dot a(t)>0$. At the bouncing point, $\dot a(t)=0$,
and around this point $\ddot a(t)>0$ for a period of time.
Equivalently, in the bouncing cosmology the hubble parameter $H$
runs across zero ($H=0$) from $H<0$ to $H>0$. In the bouncing point
we have $H=0$. A successful bounce requires that around this point
the following condition should be satisfied [30],
\begin{equation}
\dot H=-4\pi G\rho (1+w)>0~.
\end{equation}
From this relation we see that $\omega<-1$ in a neighborhood of the
bouncing point. After the bounce, the universe needs to enter into
the hot Big Bang era, otherwise the universe filled with the matter
with an equation of state parameter $\omega<-1$ will reach the big
rip singularity as what happens to the phantom dark energy [36].
This requires that the equation of state parameter of dark energy
component transits from $\omega<-1$ to $\omega>-1$. One can see from
figures (2) and (3) that in our setup a non-singular bouncing
happens with the Hubble parameter $H$ running across zero and a
minimal non-vanishing scale factor $a$. At the bouncing point
$\omega$ approaches a finite negative value. So it is possible to
realize bouncing solutions in a model universe with non-minimally
coupled dark energy component on the moving domain wall in
background AdS$_{5}$ bulk.
\begin{figure}
\begin{center}\includegraphics{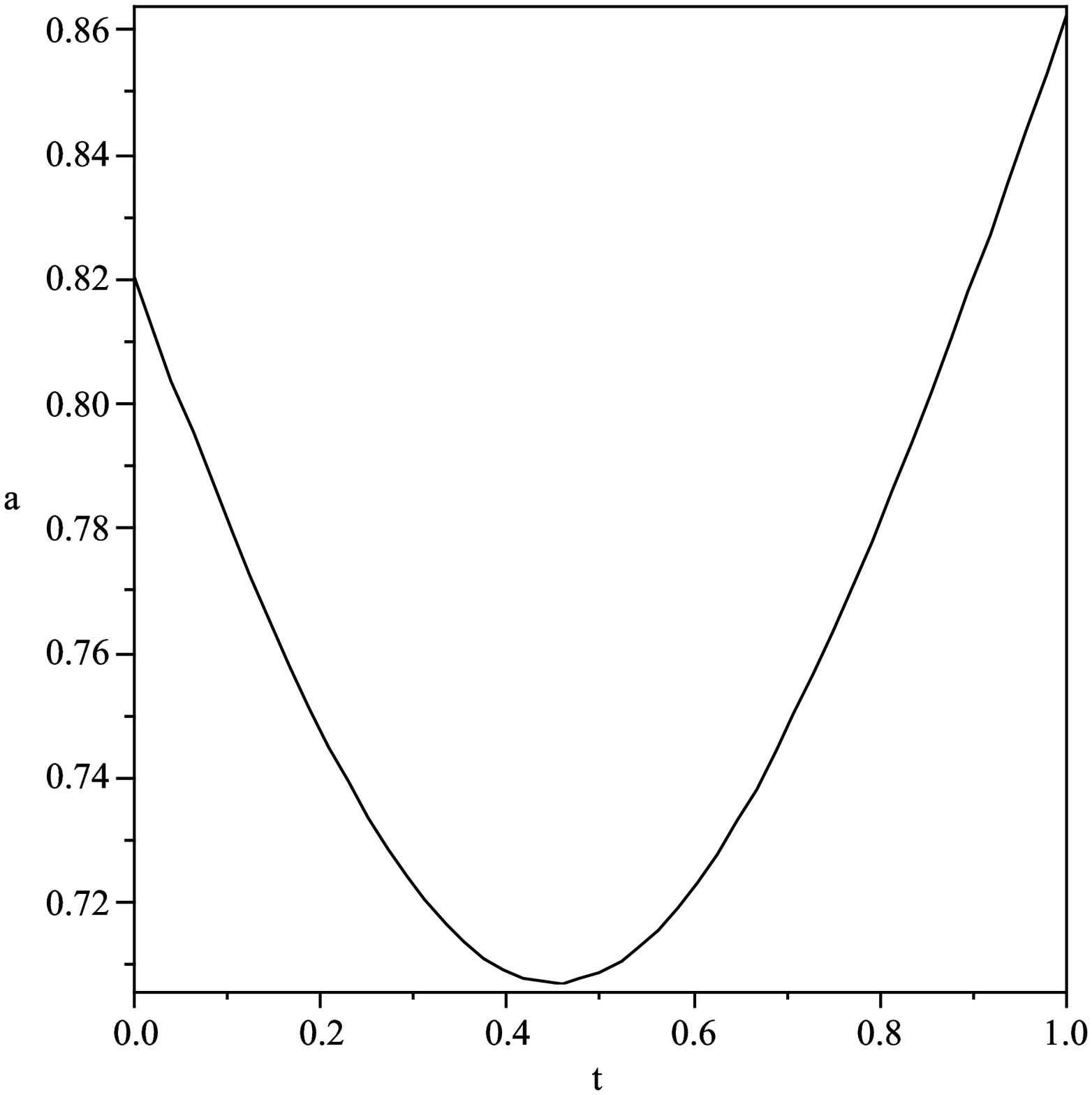} \vspace{5.5cm}\includegraphics{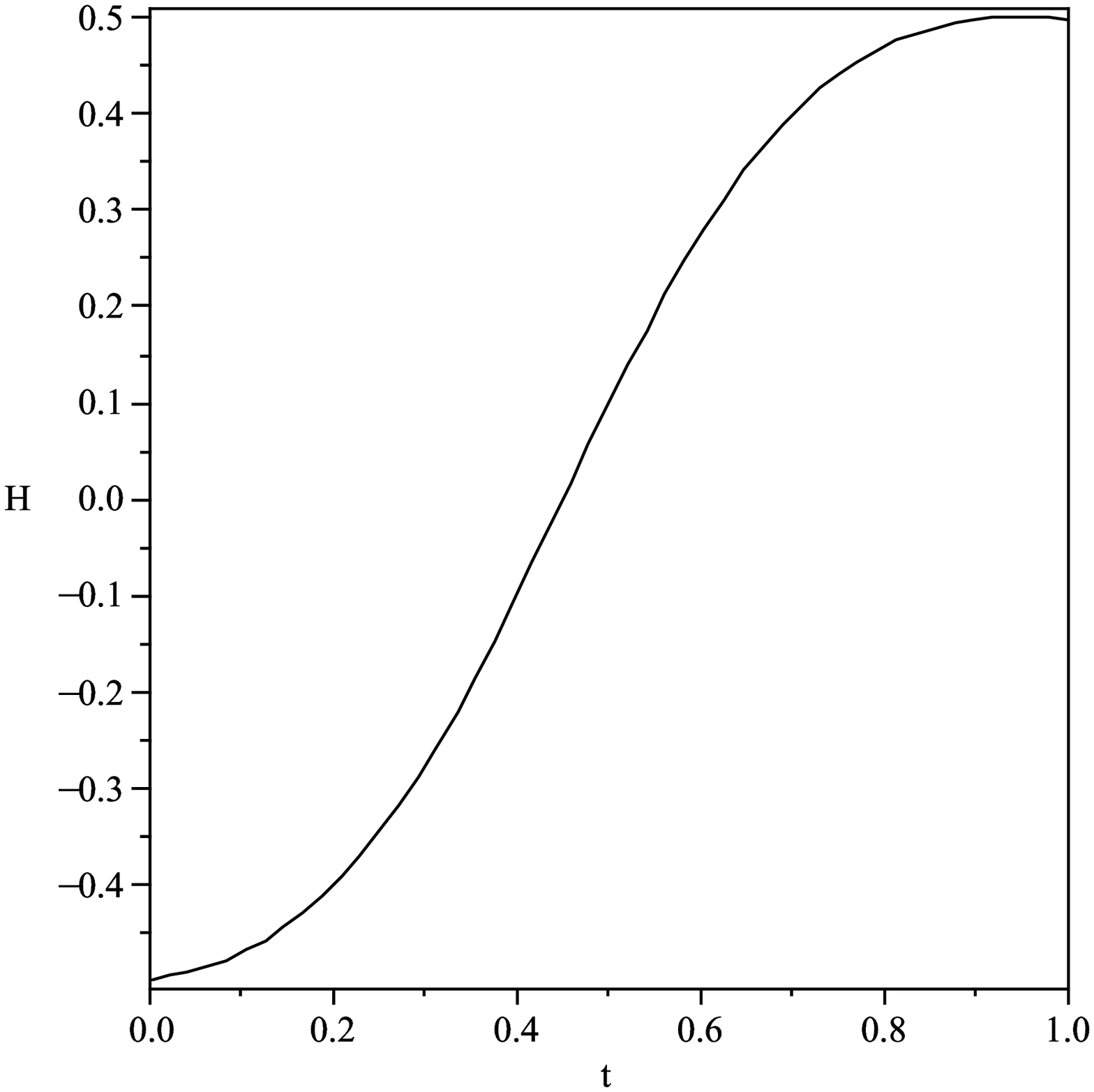}
\end{center}
\caption{\small {Variation of the scale factor $a$ relative to
cosmic time $t$ for $t_0=1$ and $\nu=-3$ (left), and $H$ (right). }}
\end{figure}

\section{Summary}
Light-curves analysis of several hundreds type Ia supernovae, WMAP
observations of the cosmic microwave background radiation and other
CMB-based experiments have shown that our universe is currently in a
period of accelerated expansion. In this respect, construction of
theoretical frameworks with potential to describe positively
accelerated expansion and crossing of the phantom divide line by the
equation of state parameter, itself is an interesting challenge.
According to existing literature on dark energy models, a {\it
minimally} coupled scalar field in $4$-dimension is not a good
candidate for dark energy model that its equation of state parameter
crosses the phantom divide line. On the other hand, a scalar field
{\it non-minimally} coupled to gravity in $4$-dimension has the
capability to be a suitable candidate for dark energy which provides
this facilities. In this paper, we have extended the nonminimal dark
energy model to a barneworld setup that brane is considered to be a
moving domain wall in a static bulk background of AdS$_{5}$ type. In
this braneworld setup, non-minimally coupled scalar field provides
even more reliable candidate for dark energy due to wider parameter
space. In fact, non-minimal coupling of the scalar field and gravity
arises at the quantum level when quantum corrections to the scalar
field theory are considered. Even if for the classical, unperturbed
theory this non-minimal coupling vanishes, it is necessary for the
renormalizability of the scalar field theory in curved spacetime
[27]. Due to complication of dynamical equations, we have restricted
our study to some specific form of non-minimal coupling ( conformal
coupling) and scalar field potentials and also we have considered
some special form of time evolution for scale factor and scalar
field via a phenomenologically reliable ansatz. Our results show
that such a model universe with nonminimally coupled scalar field as
dark energy component avoids the problem of the Big Bang
singularity. In fact, this model allows for bouncing solution which
solves the singularity problem.

The combined dataset from distant supernovae SNIa, baryon acoustic
oscillation peak and the cosmic microwave background radiation show
that the non-minimal coupling parameter $\xi$ is closed to its
conformal value of $\frac{1}{6}$, [32]. On the other hand recent
observations show that crossing of the phantom divide line occurred
at $z\simeq 0.25$, [12]. This observations can be used to confront
our model with observations. In fact, using figure $2$ and the
relation $1+z=\frac{a_{0}}{a(t)}$, we see that in our model crossing
of the phantom divide line with $\xi=0.15$ occurs at $z=0.263$ which
is close to the observationally supported value of $z\simeq 0.25$.
From another viewpoint, we can obtain a constraint on the value of
the non-minimal coupling by assuming that crossing occurs at
redshift near to $z\simeq 0.25$.

In summary, a model universe with non-minimally coupled scalar field
on the moving domain wall in AdS$_{5}$ bulk allows to explain both
early and late-time accelerated expansions. This model realizes
crossing of the phantom divide line and solves the standard Big Bang
singularity problem by admitting bouncing solutions.\\

{\bf Acknowledgment}\\
This work has been supported partially by Research Institute for
Astronomy and Astrophysics of Maragha, Iran.

\end{document}